\documentclass[a4paper,12pt]{article}
\pdfoutput=1
\usepackage{amssymb,amsfonts}
\usepackage{bbm}
\usepackage{bm}
\usepackage{mathrsfs}
\usepackage{cite}
\usepackage[scale={0.78,0.8}]{geometry}
\usepackage{multirow}
\usepackage{relsize}
\usepackage[subrefformat=parens,labelformat=parens]{subfig}
\usepackage{xcolor}
\usepackage{caption}
\usepackage{cancel}
\usepackage{graphics,graphicx}
\usepackage{subfig}
\usepackage{float}
\usepackage[intlimits,centertags]{amsmath}
\usepackage[]{hyperref}
\numberwithin{equation}{section}
\numberwithin{figure}{section}

\def\beq{\begin{align}}
\def\be{\begin{equation}}
\def\ee{\end{equation}}
\def\eeq{\end{align}}
\def\beqa{\begin{eqnarray}}
\def\eeqa{\end{eqnarray}}

\newcommand{\intf}{\int_\mathcal{M}d^4x \sqrt{-g}}
\newcommand{\intt}{\int_{\partial\mathcal{M}}d^3\xi \sqrt{-^{(3)}g}}
\newcommand{\mnr}{{\mu\nu\rho}}
\newcommand{\mnrs}{{\mu\nu\rho\sigma}}
\newcommand{\sumij}{\sum_{i=1}^J}

\begin{document}

\begin{center}
\LARGE{Environmental selection of the \\ cosmological constant \\ and\\ electroweak scales}
\\[13mm]
  \large{Matteo Moretti$^{1}$, Francisco G. Pedro$^{1,2}$ \\[6mm]}
\small{
${}^1$
Dipartimento di Fisica e Astronomia, Universit\`a di Bologna, \\ via Irnerio 46, 40126 Bologna, Italy \\ \vspace{5mm}
${}^2$
INFN, Sezione di Bologna, viale Berti Pichat 6/2, 40127 Bologna, Italy
}
\end{center}
\vspace{2cm}
\begin{abstract}
We investigate whether selfish/Goldilocks Higgs models can be extended to accomodate vacua with both the right Higgs mass and cosmological constant. Through the introduction of multiple four-form fields coupled with the Higgs scalar boson we find that both the cosmological constant and the electroweak hierarchy problems can be addressed simultaneously, without inserting the relevant physical scales by hand and avoiding fine tuning. The resulting bounds on the brane charges, the only dimensionful free parameter of the theory, depend on the strength of the four-form scalar coupling and can be of the same order of magnitude as those necessary for a solution to the cosmological constant problem alone via the Bousso-Polchinski mechanism. 
\end{abstract}
\newpage

\setcounter{page}{1}
\pagestyle{plain}
\renewcommand{\thefootnote}{\arabic{footnote}}
\setcounter{footnote}{0}

\tableofcontents

%
\section{Introduction}

Naturalness has been a powerful guiding principle in the construction of the standard model of particle physics, that has multiple times indicated the emergence of new physics at some higher energy scale before such scale could be probed experimentally. In the context of the electroweak (EW) hierarchy, naturalness is the pillar that supports the expectations of finding new physics (be it supersymmetry, compositeness, large extra dimensions or any other viable alternative) at the $TeV$ scale. As of now, no compelling sign of new physics has been found which prompts us to explore alternative solutions to the electroweak hierarchy problem that do not rely on naturalness. While it may be premature to abandon naturalness altogether, it is certainly worthwhile to explore alternatives to it even more so given the cosmological constant problem. Cosmological observations indicate that nowadays the dominant component of the universe is vacuum energy. In itself this observation poses no problem, given that quantum fields through their zero point energy will source the cosmological constant (CC). It is the fact that observations indicate that $\Lambda\approx 10^{-120} M_P^4$ that requires a justification that cannot rely on naturalness. A radically different approach to the CC problem is based on the anthropic principle that can be phrased as: observers will measure a value of $\Lambda$ that is compatible with their very existence. Such approach has been used in \cite{Weinberg:1987dv} to put an upper bound on $\Lambda$, beyond which the CC's negative pressure would prevent galaxy formation and would hence preclude the existence of intelligent life as we know it. In this context the CC would therefore be an environmental variable that can take different values in different universes, of which only a subset would be hospitable to life. One could envisage a situation whereby the same reasoning can be applied to the Higgs' mass, leading to an environmental solution to the electroweak hierarchy problem \cite{Agrawal:1997gf}.

In four dimensional field theory such environmental selection mechanism can be implemented using three-form gauge fields, their four-forum field strengths and branes charged under such gauge symmetries.

The use of free four-forms for a dynamical neutralisation of the cosmological constant has a long history, see e.g. \cite{Hawking:1984hk,Duff:1989ah,Brown:1987dd,Brown:1988kg}, culminating in the Bousso-Polchinski (BP) model \cite{Bousso:2000xa} that provides an example of a landscape encompassing a multitude of vacua and of a mechanism for transitioning between different vacua. This idea is fundamental in the context of the string landscape and anthropic arguments regarding the smallness of the observed cosmological constant.

Linearly coupled four-forms have instead been employed in an inflationary context as a way to simultaneously have a quadratic scalar potential and an unbroken shift symmetry granting it radiative stability \cite{Kaloper:2008fb,Kaloper:2011jz}. These seemingly contradictory features arise through multi-branched potentials, with each branch corresponding to a different four-form background and transition between different branches taking place via the nucleation of membranes charged under the three-form gauge field.

More recently the authors of \cite{Kaloper:2019xfj,Giudice:2019iwl} have analysed the case of a quadratically coupled scalar-four-form system, with the aim of providing a mechanism to dynamically scan the Higgs mass, an approach that builds on earlier work \cite{Dvali:2004tma} and that is in spirit akin to that of \cite{Graham:2015cka,Herraez:2016dxn}. In this paper we will review and extend this mechanism in an attempt to overcome two of its less attractive aspects, namely the fact that the electroweak scale has to be inserted by hand and that the difficulty in simultaneously having the correct CC and Higgs' mass.

This paper is structured as follows: in Sec. \ref{sec:2} we review how four-forms have been used to separately address the cosmological constant \cite{Bousso:2000xa} and electro-weak hierarchy  \cite{Kaloper:2019xfj,Giudice:2019iwl} problems; in Sec. \ref{sec:J4Forms} we review the model of  \cite{Kaloper:2019xfj,Giudice:2019iwl} and investigate whether extending it to a large number of coupled four-forms leads to vacua with both the observed cosmological constant and Higgs mass without fine tuning, analysing the constraints on flux-space and the resulting bounds on the brane charges and in Sec. \ref{sec:conclusions} we present our conclusions.

\section{Four-forms and environmental selection of the electroweak scale}\label{sec:2}

Let us consider a three-form gauge field $A_3$ and a  scalar $h$ and introduce a quadratic coupling bewteen the two. Furthermore let us assume that there are branes charged under $A_3$, the action for the system can then be written as:
\begin{equation}
\label{eq:KW action}
S = S_A + S_{h}+ S_{DJ}\ .
\end{equation}

The term $S_A $ encompasses both the three-form kinetic term and the action for a membrane of tension $T$ and charge $q$:
\begin{align}
S_A = & \intf \left[ -\frac{1}{2\cdot 4!} F_\mnrs F^\mnrs \right] +  \nonumber \\
\label{eq:three-form action}
& -T \intt + \frac{q}{6} \int_{\partial\mathcal{M}} d^3\xi \left[ A_\mnr  \frac{\partial x^\mu}{\partial \xi^a}\frac{\partial x^\nu}{\partial \xi^b}\frac{\partial x^\rho}{\partial \xi^c} \epsilon^{abc} \right].
\end{align}
where $F_\mnrs=4 \partial_{[\mu} A_{\nu \rho\sigma ]}$, $\xi$ are the coordinates adapted to the membrane $\partial\mathcal{M}$, $^{(3)}g$ is the determinant of the induced metric and $\epsilon^{abc}$ is the Levi-Civita symbol in three dimensions.

The term $S_{h}$ gathers the scalar dependent terms
\begin{equation}
\label{eq:SM action}
S_{h} = \intf \left[ -\frac{1}{2}\partial_\mu h \partial^\mu h - V \right],
\end{equation}
where $V$ includes self interactions and a quadratic coupling to the four-form field strength\footnote{In our notation $g$ denotes the four-form to scalar coupling strength while $\sqrt{-g}=(-\det(g_{\mu\nu}))^{1/2}$.}:
\begin{equation}
\label{eq:H potential}
V = -\frac{v^2}{2} h^2 + \frac{\lambda}{4} h^4 - \frac{g}{24} h^2 \epsilon_\mnrs F^\mnrs.
\end{equation}
Here $v^2$ and $\lambda$ are the bare scalar mass and quartic coupling respectively. As we'll see in this model they are not fixed to their measured values due to their couplings to the four-form.

For consistency, it is necessary to also consider the Duncan-Jensen term, denoted $S_{DJ} $ that  takes the form
\begin{equation}
\label{eq:DJ + h action}
S_{DJ} = \frac{1}{6}\int \, d^4x \, \nabla_\mu\left[ \sqrt{-g} \left( F^\mnrs A_{\nu\rho\sigma} - g\epsilon^\mnrs A_{\nu\rho\sigma}h^2 \right)\right].
\end{equation}
This is a boundary term and as such it does not contribute to the equations of motion, but it is fundamental if one wants to be able to substitute a field configuration directly into the action (and not into the field equations) as explained in \cite{Duff:1989ah}. We note that equivalently we could have included the Aurilia-Nicolai-Townsend term  to accomplish the same end result \cite{{Aurilia:1980xj}}.

Away from sources, the dynamics of $A_\mnr$, is determined by
\begin{equation}
\label{eq:Eom for F with Higgs}
\nabla_\mu \left(F^\mnrs  -g h^2 \epsilon^\mnrs \right) = 0\ ,
\end{equation}
where $\epsilon^{0123}=1$. Equation \eqref{eq:Eom for F with Higgs} admits the solution
\begin{equation}
\label{eq:Eom for F with Higgs, solved}
F^\mnrs = \left( gh^2 + c \right) \epsilon^\mnrs/ \sqrt{-g}\ ,
\end{equation}
for arbitrary constant $c$ of mass dimension two.

Substituting this result into the Duncan-Jensen action \eqref{eq:DJ + h action} we find:
\be
S_{DJ} = \frac{1}{24} \intf \left[ F_\mnrs F^\mnrs - gh^2 \epsilon_\mnrs F^\mnrs \right].
\ee
Allowing for a bare cosmological constant $\Lambda$ and putting the four-form on shell, one finds the effective action for the Higgs 
\be
S_{eff} =\intf \left[- \bar\Lambda + \frac{\bar{v}^2}{2}h^2 - \frac{\bar{\lambda}}{4}h^4 \right],
\ee
where 
\be
\label{eq:physical mass}
\bar{v}^2 = v^2 - 2cg\ , 
\ee
\be
\label{eq:physical quartic coupling}
\bar{\lambda} = \lambda + \frac{1}{2}g^2\ , 
\ee
and
\be
\label{eq:physical CC}
\bar{ \Lambda} = \Lambda + \frac{1}{2}c^2
\ee
are the \emph{physical} mass, quartic coupling and cosmological constant respectively\footnote{We adopt the notation whereby a barred parameter is the physical one.}.

From Eq. \eqref{eq:physical CC} we recover the Brown-Teitelboim (BT) scenario \cite{Brown:1988kg,Brown:1987dd}, with the physical cosmological constant $\bar{\Lambda}$ being given by the net sum of the bare $\Lambda$ and of the four-form contribution $c^2$. The presence of the quadratic coupling also implies that the physical mass $\bar{v}^2$ of the Higgs scalar is lowered by an amount proportional to the four-form magnitude $c\ (>0)$. 
 
The scanning of the physical parameters is made possible by the presence of membranes coupled to the three-form gauge field as per \eqref{eq:three-form action} with the four-form quantisation implying that the four-form background value is a multiple of the elementary brane charge, $c=n q$, which leads to 
\begin{align}
\label{eq:Higgs vev}
& \bar{v}^2 = v^2 - 2gnq\ , \\
\label{eq:Cosm conts, non-broken EW}
& \bar{\Lambda} = \Lambda + \frac{1}{2}n^2 q^2,
\end{align}
where $q$ is the membrane charge and  $n$ indicates the flux-number of a given background.

In the same way that Eq. \eqref{eq:Cosm conts, non-broken EW} leads to a dynamical neutralisation of a large and negative $\Lambda$ by means of membrane nucleation and consequent lowering of $|n|$, the same physical process can lower the Higgs mass, Eq. \eqref{eq:Higgs vev}, from its bare value $v^2$, presumably at some high energy cut off, down to the EW scale $\bar{v} \approx 246 \, \text{GeV}$.

In order to avoid fine tuning problems it is imperative that the step size between adjacent vacua, controlled by the brane charge $q$, is at most of the order of the physical scale of interest. If this mechanism is to provide an explanation for the smallness of the EW scale one must require
\begin{equation}
\label{eq:ge sim m_h ^2}
gq \lesssim m_h ^2\sim 10^{-34} M_P^2\ ,
\end{equation}
which sets the brane charge to be at or below the EW scale squared unless the coupling $g$ is extremely small. 
If on the other hand one is interested in solving the cosmological constant problem, the charges must be even smaller: 
\begin{equation}
q^2 \lesssim \bar{\Lambda}\sim10^{-120} M_P^4\ .
\end{equation}
It is obvious that in later regime the model can simultaneously address both problems.
This is a common feature of selection mechanisms based on a single four-form and has previously been discussed in the context of the cosmological constant in \cite{Bousso:2000xa}: the membrane charge must be of order of the physical scale one is trying to explain and therefore hierarchically below the UV cut off of the theory. In the context of the BT mechanism this translates into having brane charges of order of the observed cosmological constant and in the selfish/Goldilocks Higgs model the charges are constrained to be around the EW scale. Why should these extended objects have such small charges is puzzling as one would expect fundamental branes to have charges of order one in units of the fundamental scale of the theory.

Focusing on the regime of Eq. \eqref{eq:ge sim m_h ^2} we note that inside each bubble with $\bar{v}^2<0$ the  electroweak symmetry can be  spontaneously broken. The flux number of the last nucleation $n_*$ is given by
\begin{equation}
\label{eq:n_* last nucleation}
n_* = \left[ \frac{v^2}{2g q} \right] \ ,
\end{equation}
where $[x]$ denotes the integer part of $x$. The number of nucleations is forced to be exactly equal to $n_*$ since:
\begin{itemize}
\item
If $n > n_*$: in this case the value of $\bar{v}^2$ is too high (and negative) making the EW symmetry broken at a higher scale energy, as a consequence all particles would be heavier than measured;

\item
If $n < n_*$: in this case the EW symmetry would not be broken and the effective theory would contain only massless particles.
\end{itemize}

Given that in the true vacuum of the Higgs potential $\langle h \rangle=\bar{v}/\sqrt{\bar{\lambda}}$
\be
\langle V \rangle=-\frac{1}{4}\frac{\bar{v}^2}{\bar{\lambda}}
\ee
it follows that in the broken phase, the physical cosmological constant becomes
\begin{equation}
\label{eq:Cosm const, broken EW}
\bar{\Lambda} = \Lambda + \frac{1}{2}n^2 q^2 - \frac{1}{4}\frac{(v^2 - 2gnq)^2}{\bar\lambda}.
\end{equation}

With the brane charge fixed by the requirement of having the correct EW scale, the typical jump in the physical cosmological constant between adjacent vacua (in flux space) is of order
\begin{equation}
\label{eq:Cosm const last nucleation}
\Delta\Lambda \big{|}_{n_*} \sim \frac{v^2 q}{g}\ ,
\end{equation}
a value many orders of magnitude above the observed value. This forces us to assume a high degree of tuning in order to reach the actual value of the cosmological constant, or to accept that in this landscape, universes with the right Higgs mass and cosmological constant are very rare indeed.\\
What we try to do in the next section is to modify this latter mechanism by including a larger number $J$ of different four-forms in order to avoid the gap problem, in the same spirit of the BP extension of the BT mechanism \cite{Bousso:2000xa}.


\section{Environmental selection of the electroweak scale: multiple four-forms}
\label{sec:J4Forms}

Let us now extend the mechanism of \cite{Kaloper:2019xfj,Giudice:2019iwl} by considering an arbitrary number $J$ of four-forms coupled to the scalar. The action takes the form
\begin{multline}
\label{eq:New interesting action}
S = \int_\mathcal{M} d^4x \sqrt{-g} \left[ -\Lambda + \frac{1}{2}(\partial h)^2 +\frac{v^2}{2}h^2 -\frac{\lambda}{4}h^4 + \sumij \left( \frac{g_i}{24}h^2\epsilon_\mnrs F_{(i)}^\mnrs - \frac{1}{48}F_{(i)\mnrs} F_{(i)}^\mnrs \right) \right] + \\
+ \frac{1}{6} \int_\mathcal{M} d^4x \nabla_\mu \left[ \sqrt{-g} \sumij \left( F_{(i)}^\mnrs A_{(i)\nu\rho\sigma} - g_i \epsilon^\mnrs A_{(i)\nu\rho\sigma} h^2 \right) \right].
\end{multline}
Then we look for the Euler-Lagrange equations for the four-form fields, which, given the structure of the \eqref{eq:New interesting action}, are a mere generalization of the $J=1$ case and admit the solution
\begin{equation}
\label{eq:New four-forms eqs of motion}
F_{(i)}^\mnrs = \left( g_i h^2 + c_i \right)\epsilon^\mnrs / \sqrt{-g}\ ,
\end{equation}
where $c_i$ are integration constants. Substituting these back into the action \eqref{eq:New interesting action} and taking into account the Duncan-Jensen term, we find the effective action for the scalar:
\be
S  = \int_\mathcal{M} d^4x \sqrt{-g} \left[ -\bar\Lambda + \frac{\bar{v}^2}{2}h^2 - \frac{\bar{\lambda}}{4}h^4 \right],
\label{eq:New solved interesting action}
\ee
where $\bar{v}^2 = v^2 - 2\sumij c_i g_i$, $\bar{\lambda} = \lambda + 2\sumij g_i^2$ and $\bar{\Lambda} = \Lambda + \frac{1}{2} \sumij c_i^2$ .
\\Using the Dirac quantization condition $c_i = n_i q_i$ one finds:
\begin{align}
\label{eq:New true vev}
& \bar{v}^2 = v^2 - 2\sumij g_i n_i q_i\ , \\
\label{eq:New true self coupling}
& \bar{\lambda} = \lambda + 2\sumij g_i^2\ , \\
\label{eq:New true cosm const}
& \bar{\Lambda} = \Lambda + \frac{1}{2}\sumij n_i^2 q_i^2\ , 
\end{align}
a direct generalisation of Eqs. \eqref{eq:physical mass},\eqref{eq:physical quartic coupling} and \eqref{eq:physical CC}. It is worth noting at this point that unlike $\bar \Lambda$ and $\bar v^2$ that depend on the flux integers $n_i$, the Higgs quartic coupling is independent of $n_i$ and therefore does not scan in the flux landscape.

Electroweak symmetry breaking will not be possible everywhere on flux-space, in fact depending on the factor $\sumij g_i n_i q_i$ one can have, in complete analogy with the single four-form case
\begin{equation}
  \bar{v}^2 = \begin{cases}
    > 0 & \Rightarrow \text{No EWSB} \Rightarrow \bar{\Lambda} = \Lambda+ \frac{1}{2}\sumij n_i^2 q_i^2 \\
    \leq 0 & \Rightarrow \text{EWSB} \Rightarrow \bar{\Lambda} = \Lambda + \frac{1}{2}\sumij n_i^2 q_i^2 - \frac{1}{4}\frac{\left( v^2 -2\sumij g_i n_i q_i \right) ^2}{\bar{\lambda}}
  \end{cases}.
\end{equation}
In the second and most interesting case the electroweak symmetry is broken and thus the scalar sector gives a contribution to the cosmological constant. In such a configuration, for the last nucleation the value of the cosmological constant is:
\begin{equation}
0 \, < \, \Lambda + \frac{1}{2}\sumij n_i^2 q_i^2 - \frac{\left(v^2 -2\sumij g_i n_i q_i \right)^2}{4\bar{\lambda}} \, <  \Delta\Lambda,
\end{equation}
where $\Delta \Lambda$ is the observed value of the cosmological constant. Assuming for concreteness $\Lambda<0$ this can be rewritten as \footnote{ Note that the scalar-four-form coupling allows for the extension of the BP mechanism to $\Lambda>0$. In such case the above results are easily modified by substituting $| \Lambda|\rightarrow -\Lambda$  and so we see that $R \in \mathbb{R}$ when $\Lambda< \frac{v^4}{4\bar{\lambda}}\left( 1 + \frac{2 J g^2}{\bar{\lambda}} \right)$, which allows for positive bare CC for a $\frac{\Lambda}{v^4}<\frac{1}{4\bar{\lambda}}\left( 1 + \frac{2 J g^2}{\bar{\lambda}} \right)$. We'll show below that a simultaneous solution to the CC and EW problems also provides a lower bound on the quantity $\frac{\Lambda}{v^4}$. }:
\begin{equation}
\label{eq:New bound for the cosmological constant}
2|\Lambda| \, < \, \sumij n_i^2 q_i^2 - \frac{\left(v^2 -2\sumij g_i n_i q_i \right)^2}{2\bar{\lambda}} \, < \, 2\left( |\Lambda| + \Delta\Lambda \right).
\end{equation}

In order to simplify the analysis we henceforth assume $g_i=g\ \forall\ i$,  this allows us to rewrite \eqref{eq:New bound for the cosmological constant} as 
\begin{equation}
2 |\Lambda| \, < \, \,\left(1-\frac{2 g^2}{\bar{\lambda}}\right) \sumij \left( n_i q_i\right)^2 -\frac{v^4}{2\bar{\lambda}}+ \frac{2 v^2 g}{\bar{\lambda}}  \sum_{i=1}^J n_i q_i-\frac{2g^2}{\bar{\lambda}} \sum_{i=1}^J n_i q_i \sum_{j=1 (j\neq i)}^{J} n_j q_j < 2( |\Lambda|+\Delta \Lambda)\ .
\end{equation}
For  $\frac{2g^2}{\bar{\lambda}} \in \ ]0,1[$, a natural range for this parameter combination, these relations define the volume contained within a pair of $J$ dimensional ellipsoids. For sufficiently weak coupling $\frac{2g^2}{\bar{\lambda}}\ll1$ we may approximate the bound by
\begin{equation}
\label{eq:New h-s final bound R}
R^2 \, \lesssim \, \, \sumij \left( n_i q_i + \frac{g v^2}{\bar{\lambda}} \right)^2 \lesssim \left( R + \Delta R \right) ^2,
\end{equation}
where we have defined
\begin{align}
\label{eq:New R}
& R = \sqrt{2 \left[ |\Lambda| + \frac{v^4}{4\bar{\lambda}} \left( 1 + \frac{2 J g^2}{\bar{\lambda}} \right) \right]}\ , \\
\label{eq:New RDR}
& R + \Delta R = \sqrt{2 \left[ |\Lambda| + \frac{v^4}{4\bar{\lambda}} \left( 1 + \frac{2 J g^2}{\bar{\lambda}} \right) + \Delta\Lambda \right]}\ .
\end{align} Written in this manner we promptly identify this constraint as a spherical shell of inner radius $R$ and thickness  $\Delta R = \frac{\Delta\Lambda}{R}+\mathcal{O}(\Delta\Lambda^2)$ embedded in a $J$-dimensional space.

The presence of scalar field and its coupling to the four-forms brings about a couple of differences with respect to the standard BP mechanism, in particular:
\begin{itemize}
\item the radius $R$ and thickness of the shell $\Delta R$ now depend also on $g, v$ and $J$;
\item the center of the sphere is displaced from the origin.
\end{itemize}
If we focus exclusively on the cosmological constant problem \cite{Bousso:2000xa}, demanding that this mechanism leads to solutions compatible with the observed value leads to a bound on the brane charges. In particular, solutions exist provided the volume of the shell
\be
\mathcal{V}_{shell} \approx \frac{2 \pi^{J/2}}{\Gamma(J/2)} R^{J-1} \Delta R
\label{eq:Vshell}
\ee
is greater than that of the unit cell in flux space $\mathcal{V}_{cell}=q^J$, implying 
\be
q\le \frac{2^{1/J} \pi^{1/2}}{\Gamma(J/2)^{1/J}} R^{1-1/J} \Delta R^{1/J}\ .
\ee
Assuming that the bare values for the CC and Higgs mass lie at the same UV scale $|\Lambda|=v^4=M_{UV}^4$ one may approximate $R\approx \mathcal{O}(1) M_{UV}^2$ and $\Delta R \approx \frac{\Delta \Lambda}{R}$ which yields
\be
q\le\mathcal{O}(1) \Delta \Lambda^\frac{1}{J}(M_{UV})^{2-4/J}\equiv q_{BP}\ ,
\label{eq:qCCBound}
\ee
up to $\mathcal{O}(1)$ factors, in agreement with  the estimates of \cite{Bousso:2000xa}.

Let us now turn our attention to the EW sector. Unlike for the cosmological constant, where the constraints on flux space are quadratic and therefore have spherical symmetry, the constraints on the Higgs vev are linear and take the form
\begin{equation}
\label{eq:New2 hyperplanes}
0 \, \lesssim \, v^2 - 2\sumij g_i n_i q_i \, \lesssim \, \Delta v^2  \ .
\end{equation}
These inequalities define a region of flux-space contained between two hyperplanes, whose coordinates are $x_i := n_i q_i$. 
If for simplicity one assumes that $g_i=g\ \forall\  i$,  Eq. \eqref{eq:New2 hyperplanes} reduces to
\be
\frac{v^2 - \Delta v^2}{2g} \, \lesssim \, \sumij n_i q_i \, \lesssim \, \frac{v^2}{2g}\ .
\label{eq:constraintV}
\ee
It is worth noting that the volume of this region is infinite and therefore the simple two dimensional extension of the mechanism of \cite{Kaloper:2019xfj,Giudice:2019iwl} generates vacua with the correct EW scale, regardless of the value of the charges. Clearly, unless $q$ is suitably small (c.f. Eq.\eqref{eq:qCCBound}) such vacua will have an unacceptably large $\bar{\Lambda}$.
If vacua with both the right $\bar \Lambda$ and $\bar v^2$ are to exist one must demand that the volume of the region defined by the constraints \eqref{eq:New h-s final bound R} and \eqref{eq:constraintV}, $\mathcal{V}_\cap$  obeys
\be
\mathcal{V}_\cap\ge \mathcal{V}_{cell},
\ee
which translates into a new, stricter upper bound on the brane charges. Note that the geometry of the constraints also depends on the value of the coupling $g$ and that a simultaneous solution of both constraints only takes place for certain values of the coupling, as we will now demonstrate.

\subsection{Constraints on the Higgs sector}

The Higgs potential, at experimentally probed scales is fixed by the measurement of the W and Z gauge boson masses which fix its vev and its own mass $m_H=125 \text{GeV}$ which fixes $\bar{v}$. These bounds can be translated into \cite{ParticleDataGroup:2020ssz}
\be
\bar{\lambda}\approx 0.13 \qquad\text{and}\qquad  \bar{v}\approx 246\ \text{ GeV}\ .
\ee

Noting that under the democratic assumption for the couplings Eq. \eqref{eq:New true self coupling} reduces to 
\be
\bar{\lambda}=\lambda+2 J g^2=\lambda+\xi\ ,
\ee
where for convenience we defined the quantity
\be
\xi\equiv 2 J g^2
\ee
as this combination of parameters plays a fundamental role in the geometry of the intersections in flux-space. Given the experimental bounds on $\bar{\lambda}$, the bare coupling $\lambda$ and the four-form contribution are constrained  by $\lambda \approx -\xi + 0.13$.
In principle one can contemplate two limiting scenarios under which this equality is satisfied:
\begin{itemize}
\item $\xi \ll \mathcal{O}(1)$: in this case the bare and physical quartic couplings are of the same order and sign: $\bar{\lambda}\approx\lambda$
\item $\xi\gg \mathcal{O}(1)$: in this case the bare coupling can be negative with the physical coupling being pushed to the measured value by  the large four-form contribution. Note that $\lambda<0$ does not prevent symmetry breaking as it is the physical coupling $\bar{\lambda}$ that determines whether or not it takes place. A precise cancellation between the bare and four-form contributions must take place in this regime, making it challenging from a naturalness point of view.  Furthermore, depending on the CP nature of $\xi$, this regime can lead to CP-violation in the Higgs sector, as discussed in \cite{Kaloper:2019xfj}.
\end{itemize}

\subsection{A bound on the Higgs-four-form coupling}

A minimum bound for the coupling $g$ can be obtained by demanding  that the spherical shell and the planes intersect. The smallest distance between the planes and the origin in $J$ dimensions can be shown to be
\be
\label{eq:New2 J distance plane-origin}
P_J(g) = \frac{v^2}{2\sqrt{J}g}\ .
\ee
Note in particular that it diverges as $g\rightarrow 0$. The center of the spherical shell is located at 
\be
\label{eq:New2 J distance center-origin}
C_J(g) = \sqrt{J} \frac{g v^2}{\bar\lambda}
\ee
and the inner radius is given by \eqref{eq:New R} which we rewrite as
\begin{equation}
\label{eq:New2 J R}
R_J(g) = \sqrt{2 \left[ |\Lambda| + \frac{v^4}{4\bar{\lambda}} \left( 1 + \frac{2 J g^2}{\bar\lambda} \right) \right]}\ .
\end{equation}
As $g\rightarrow0$ (while keeping ${\bar\lambda}$ fixed) we see that $C_J(g)\rightarrow 0$ and that $R_J(g)$ remains finite.  Note that the center of the spherical shell lies along the direction $(-1,...,-1)$ while the point on the planes that is closest to the origin lies along the $(1,1,...,1)$ direction in the $J$ dimensional flux space. It therefore emerges that a simultaneous solution to the EW and CC problems can be translated into the condition
\be
P_J+C_J\in[0, R_J]\ .
\ee
The upper limit corresponds to intersection at the north pole of the sphere while the lower corresponds to intersection along the sphere's equator.

From the condition $P_J+C_J=R_J$ one finds that the minimum value of the coupling $g$ (for a given $J$) that allows for intersection at the north pole corresponds to
\be
\label{eq:xi_min}
\xi_{\min}=\frac{\bar{\lambda}}{-1+4 \alpha^4 \bar{\lambda}}  ,
\ee
where 
\be
\alpha\equiv\frac{|\Lambda|^{1/4}}{v}
\ee
parametrises the hierarchy between the bare CC and EW scales. From its definition $\xi>0$ and therefore a simultaneous solution is only possible for 
\be
\alpha > \left(\frac{1}{4 \bar{\lambda}}\right)^{1/4}\approx 1.18 \ ,
\ee
or equivalently this mechanism cannot simultaneously address the EW and CC problems if the bare EW scale lies above the bare CC scale. This prompts us to identify the highest energy scale in the problem with that of the cosmological constant, $\Lambda^\frac{1}{4}=M_{UV}$, while $v=\frac{M_{UV}}{\alpha}< M_{UV}$.

Given the different scaling of $P_J$ and $C_J$ with $\xi$, the function $P_J+C_J$ features a minimum at the critical value
 \be
 \xi_\text{c}=\bar{\lambda}\ ,
 \ee
 where $ P_J+C_J\big | _{\xi_c}=v^2 \sqrt{2/\bar{\lambda}}$. This corresponds to the situation where the planes are at the maximal distance from the north pole of the sphere. This implies that
 \be
\frac{ P_J+C_J}{R_J}\big | _{\xi_c}=\frac{1}{\sqrt{\alpha^4 \bar{\lambda}+1/2}}\ .
 \ee
An interesting limit of this result corresponds to intersection along the equator of the $J$ dimensional sphere in flux space. This corresponds to $P_J+C_J\ll R_J$ or equivalently $\alpha^4 \bar{\lambda}\gg 1/2$.

\begin{figure}
\centering
	\includegraphics[width=0.9\textwidth]{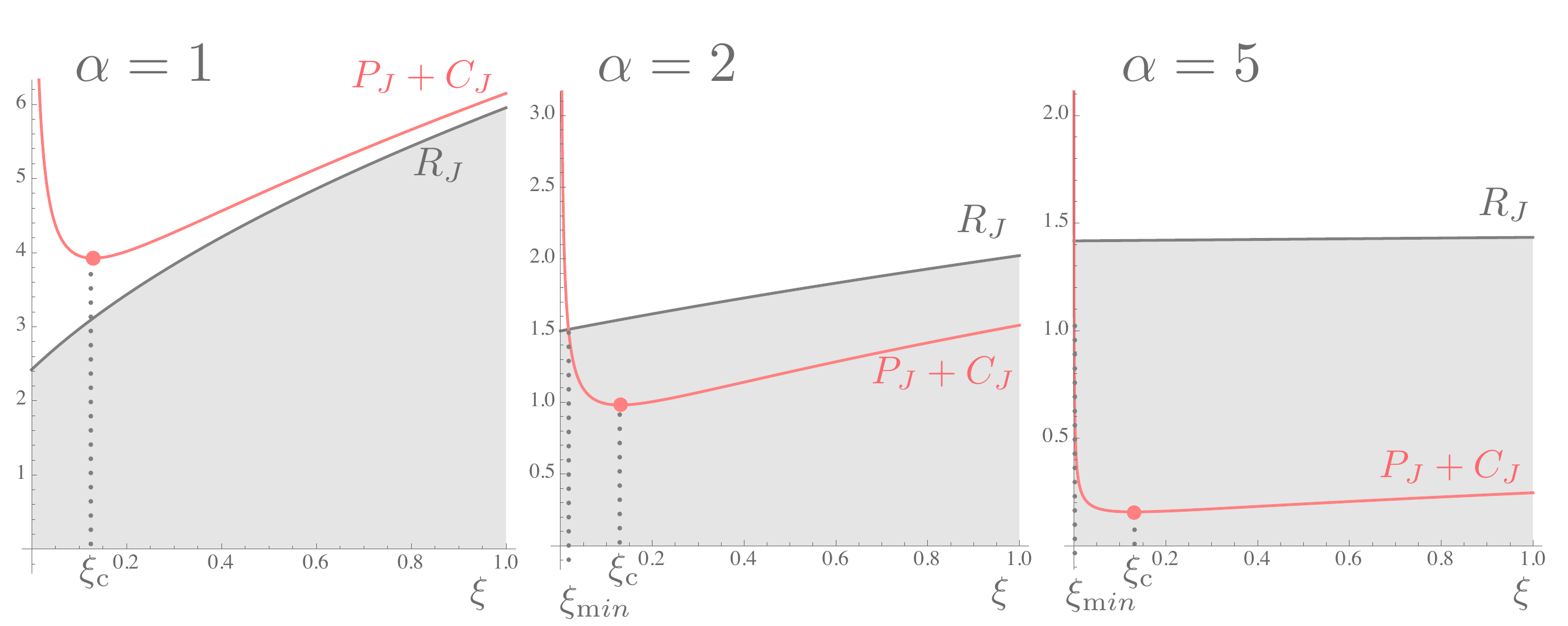}
\caption{Relative position between the planes and the spheres for different values of $\alpha\equiv\frac{|\Lambda|^{1/4}}{v}$. Simultaneous solutions to the cosmological and hierarchy problems are possible for $\xi\ge \xi_{\min}$.}
\label{fig:xiFig}
\end{figure}

For $\xi > \xi_\text{c}$ the planes approach  once again the north pole of the sphere, whose radius also grows with $\xi$. This behaviour is illustrated in Fig. \ref{fig:xiFig} for three representative values of the hierarchy $\frac{|\Lambda|^{1/4}}{v}$.

We note also that the distance between the planes decreases as they move towards the center of the sphere, since
\be
P(v^2)-P(v^2-\Delta v^2)= \frac{\Delta v^2}{\sqrt{2 \xi}}
\ee
taking the maximum value at $\xi = \xi_{min}$.

\subsection{A bound on the brane charges}
\label{sec:Vint}

In this section we analytically estimate the intersection volume $\mathcal{V}_\cap$ for arbitrary $J$ and use it to derive an upper bound of the brane charges, below which this mechanism can accomodate the observed/measured values of both the CC and EW scales. We then validate the analytical estimates by performing a numerical calculation using a Montecarlo method, whose details are presented in the Appendix.

\subsection*{Case J = 3}

\begin{figure}
\centering
	\includegraphics[width=0.8\textwidth]{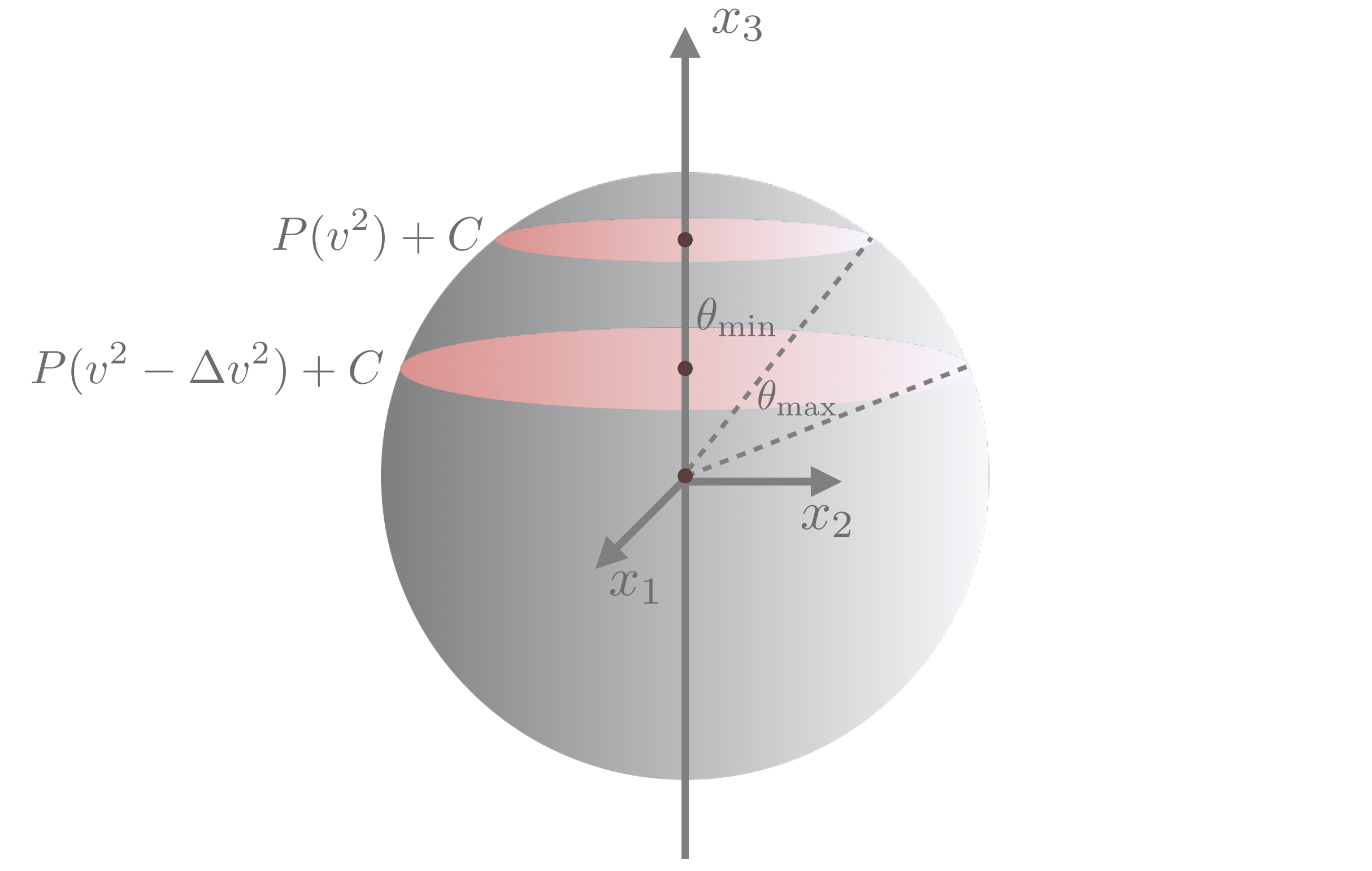}
\caption{Geometry of flux-space constraints for $J=3$. This very same configuration can be applied for larger $J$ if one replaces $x_3$ with $x_J$.}
\label{fig:3D-shell-area}
\end{figure}

We start by computing the intersection volume for the $J=3$ case before generalising the results for arbitrary $J$. Aligning the axis such that the planes are surfaces of constant $x_3$ and adopting spherical coordinates $(r,\theta,\phi)$ whereby $x_3= r \cos \theta$ we are to compute
\be
\mathcal{V}_\cap=\int_R^{R+\Delta R} r^2 dr \int_0^{2\pi} d\phi \int_{\theta_{\min}}^{\theta_{\max}} \sin \theta d\theta\ ,
\label{eq:Vint3D}
\ee
where the integration limits are
\be
\cos \theta_{\min}= \frac{P(v^2)+C}{R}
\label{eq:limInf}
\ee
and
\be
\cos \theta_{\max}= \frac{P(v^2-\Delta v^2)+C}{R}=\cos \theta_{\min}-\frac{\Delta v^2}{\sqrt{2 \xi} R}
\label{eq:limSup}
\ee
with $P, C$ and $R$ are given by \eqref{eq:New2 J distance plane-origin}-\eqref{eq:New2 J R} with $J=3$. The geometry is illustrated in Fig. \ref{fig:3D-shell-area}. Integrating Eq. \eqref{eq:Vint3D} one finds
\be
\mathcal{V}_\cap=\frac{(R+\Delta R)^3-R^3}{3}2 \pi \frac{\Delta v^2}{\sqrt{2 \xi}R}\ .
\ee
Exploiting the hierarchy $\Delta R\ll R$ and using $\Delta R\approx \Delta \Lambda/R$ one may approximate
\be
\mathcal{V}_\cap\approx \Delta \Lambda \Delta v^2 \pi \sqrt{2/\xi}\ ,
\ee
to leading order in $\Delta R/R$. We observe that for $J=3$ the volume of the intersection region is independent of the radius of the shell and that it is maximal for intersection close to the north pole, $\xi\sim\xi_{\min}$, two features that are particular to the current case and that do not survive in higher dimensions.

\subsection*{Arbitrary J}

For an arbitrary number of dimensions $J$, the procedure is identical, the only difference being the increased number of angular variables. Adopting $J$ dimensional spherical coordinates $(r, \theta_1, \theta_2, ... , \theta_{J-2},\phi)$ and aligning the axis such that the planes are surfaces of constant $x_n=r \cos \theta_1$ we may write

\be
\mathcal{V}_\cap=\int_R^{R+\Delta R} r^{J-1} dr \int_0^{2\pi} d\phi \int_{\theta_{\min}}^{\theta_{\max}} (\sin \theta_1)^{J-2} d\theta_1\prod_{i=2}^{J-2} \int_{0}^{\pi} (\sin \theta_i) ^{J-1-i}d\theta_i\ ,
\ee
leading to
\be
\mathcal{V}_\cap=\frac{(R+\Delta R)^J-R^J}{J} \frac{2 \ \pi^{\frac{J-1}{2}} }{\Gamma(\frac{J-1}{2}) } \int_{\theta_{\min}}^{\theta_{\max}} (\sin \theta_1)^{J-2} d\theta_1\ .
\ee
Let us define the remaining angular integration 
\be
\mathcal{I}\equiv\int_{\theta_{\min}}^{\theta_{\max}} (\sin \theta_1)^{J-2} d\theta_1=\int_{u(\theta_{\min})}^{u(\theta_{\max})}  (1-u^2)^\frac{J-3}{2} du\ ,
\label{eq:Integral}
\ee
where the new integration variable $u$ is given by  $u\equiv\cos \theta$ and the integration limits are given by Eqs. \eqref{eq:limInf},\eqref{eq:limSup} which we may write explicitly in terms of $\xi$ as 
\be
u(\theta_{\min})=\frac{\frac{1}{\sqrt{\xi}}+\frac{\sqrt{\xi}}{\bar{\lambda}}}{2\sqrt{\frac{|\Lambda|}{v^4}+\frac{1+\xi/\bar{\lambda}}{4\bar{\lambda}}}}
\label{eq:umin}
\ee
and
\be
u(\theta_{\max})=u(\theta_{\min}) -  \frac{\Delta v^2/v^2}{2\sqrt{\xi}\sqrt{\frac{|\Lambda|}{v^4}+\frac{1+\xi/\bar{\lambda}}{4\bar{\lambda}}}}\equiv u(\theta_{\min})-\Delta u\ .
\label{eq:umax}
\ee

The integral \eqref{eq:Integral} can be written in closed form in terms of hypergeometric functions though it is more useful to give approximate results for a few interesting cases corresponding to intersection close to the poles and along the equator.

For intersection close to the north pole, defined by $ \xi = \xi_{\min}\Rightarrow P+C=R $ one finds $u(\theta_{\min})=1$. In this regime we may approximate 
\be
\mathcal{I}(\xi_{\min})\approx \frac{2^\frac{J-1}{2}}{J-1} \Delta u^\frac{J-1}{2}=\frac{1}{J-1}\left( \frac{\Delta v^2}{v^2}\frac{4 \alpha^4 \bar{\lambda}-1}{2 \alpha^4 \bar{\lambda}}\right)^\frac{J-1}{2}
\label{eq:Ipole}
\ee
yielding
\be
\mathcal{V}_\cap(\xi_{\min}) \approx \Delta \Lambda\  (\Delta v^2)^\frac{J-1}{2}\ R^\frac{J-3}{2}  \ \xi_{\min}^\frac{1-J}{4} \ \frac{2^\frac{J+3}{4}}{J-1}\frac{\pi^\frac{J-1}{2}}{\Gamma(\frac{J-1}{2})} \ ,
\label{eq:VintPole}
\ee
where $R=R(\xi_{\min},\alpha)\in\  ] \sqrt{2},\infty [\ M_{UV}^2$, with the lower limit corresponding to large $\alpha$.

Away from the north pole we may exploit the fact that $\Delta v^2/v^2\ll1$ to approximate 
\be
\mathcal{I}(\xi)\approx (1-u(\xi)^2)^\frac{J-3}{2} \Delta u(\xi)\ ,
\label{eq:Iapprox}
\ee
which using Eqs. \eqref{eq:xi_min}, \eqref{eq:umin} and \eqref{eq:umax} can be written as
\be
\mathcal{I}(\xi)\approx \frac{\Delta v^2}{v^2}\frac{ \bar{\lambda}^{J-2} (\xi -\xi_{\min})^{\frac{J-3}{2}}  \sqrt{\xi_{\min}}}{  \left(\xi  \left(\bar{\lambda}^2+2\bar{\lambda} \xi_{\min}+\xi  \xi_{\min}\right)\right)^{\frac{J}{2}-1}}.
\ee
As $\xi \rightarrow \xi_{\min}$ the angular integral tends to zero as $(\xi-\xi_{\min})\frac{J-3}{2}$, on the other hand for $\xi\gg\bar{\lambda},\xi_{\min}$ we find that $\mathcal{I}\propto \xi^{\frac{-J+1}{2}}$ and therefore, by continuity, $\mathcal{I}(\xi)$ must have a maximum at finite $\xi$.
One can show that the maximum is located at 
\be
\xi_M=\frac{(2 J-3) \bar{\lambda}-16 \alpha ^8 \bar{\lambda}^3+\bar{\lambda}\Sigma}{2 (J-1) \left(4 \alpha ^4 \bar{\lambda}-1\right)} \ ,
\ee
and takes the value
\be
\begin{split}
\mathcal{I}(\xi_\text{M})\approx\sqrt{2}\frac{\Delta v^2}{v^2} \frac{(J-1)^{\frac{J-1}{2}}}{(J-2)^{\frac{J}{2}-1}} &(4 (\alpha^4 \bar{\lambda})-1)^{J-2} \left(\Sigma -16 (\alpha^4 \bar{\lambda})^2-1\right)^{\frac{J-3}{2}} \times \\
&\left[16 (\alpha^4 \bar{\lambda})^2 \left(4 J+\Sigma -16 (\alpha^4 \bar{\lambda})^2-6\right)+\Sigma -1\right]^{1-\frac{J}{2}}
\end{split}
\ee
where we defined $\Sigma\equiv\sqrt{32 \left(2 J^2-8 J+7\right) (\alpha^4 \bar{\lambda})^2+256 (\alpha^4 \bar{\lambda})^4+1}$. If we denote by $\mathcal{I}_{\text{tot}}$ the full angular integral for $\theta=[0,\pi]$, c.f Eq. \eqref{eq:Integral},  at large $\alpha^4 \bar{\lambda}$ one may approximate
\be
\mathcal{I}(\xi_\text{M},\alpha^4 \bar{\lambda}\gg1)\approx\frac{\Delta v^2}{v^2}\frac{(J-3)^{\frac{J-3}{2}}}{ (J-2)^{\frac{J-2}{2}} }\rightarrow^{J\gg1}\rightarrow \frac{\Delta v^2}{v^2} \sqrt{\frac{2}{\pi e}}\mathcal{I}_\text{tot}\ .
\ee
This implies that for $J\gg 1$
\be
\mathcal{V}_\cap(\xi_\text{M},\alpha^4 \bar{\lambda}\gg1)\approx\frac{\Delta v^2}{2 v^2}\mathcal{V}_{shell}
\ee
where $\mathcal{V}_{shell}$ is the volume of the spherical shell in the BP mechanism given by Eq. \eqref{eq:Vshell} and we have approximated $\sqrt{\frac{2}{\pi e}} \approx \frac{1}{2}$. This corresponds to the most favourable setup, where the volume of the viable region where a simultaneous solution to the EW and CC problems is possible is suppressed by a factor of $\frac{\Delta v^2}{2 v^2}\ll1$ with respect to the volume of the region where only the CC problem is addressed.

In the configuration where the planes are the furthest from the north pole, $\xi=\xi_c=\bar{\lambda}$, the angular integral is
\be
\mathcal{I}(\xi_\text{c})\approx\frac{\Delta v^2}{v^2} \left(-2+4 \alpha ^4 \bar{\lambda}\right)^{\frac{J-3}{2}} \left(2+4 \alpha ^4 \bar{\lambda}\right)^{\frac{2-J}{2}}\ .
\label{eq:Ixic}
\ee
If $\alpha^4 \bar{\lambda}\gg 1$ the intersection can take place close to the equator $\frac{P+C}{R}\ll1$ which corresponds to $u(\theta_{\max})=0$, one may further approximate
\be 
\mathcal{I}(\xi_\text{c},\alpha^4 \bar{\lambda}\gg1)\approx \Delta u=\frac{\Delta v^2/v^2}{2\sqrt{\bar{\lambda}}\alpha^2 }
\label{eq:Iequator}
\ee 
which yields the intersection volume
\be
\mathcal{V}_\cap(\xi_\text{c},\alpha^4 \bar{\lambda}\gg1)\approx \Delta \Lambda\ \Delta v^2 \ \frac{R^{J-3}}{\sqrt{\xi_\text{c}}}\ \frac{\sqrt{2} \pi^\frac{J-1}{2} }{\Gamma(\frac{J-1}{2})}\ ,
\label{eq:VintEquator}
\ee
where $R=R(\xi_\text{c},\alpha)\in\  ]\sqrt{2},\sqrt{6}[ \ M_{UV}^2$ and $\xi_\text{c}=\bar{\lambda}=0.13$.

\begin{figure}
\centering
	\includegraphics[width=\textwidth]{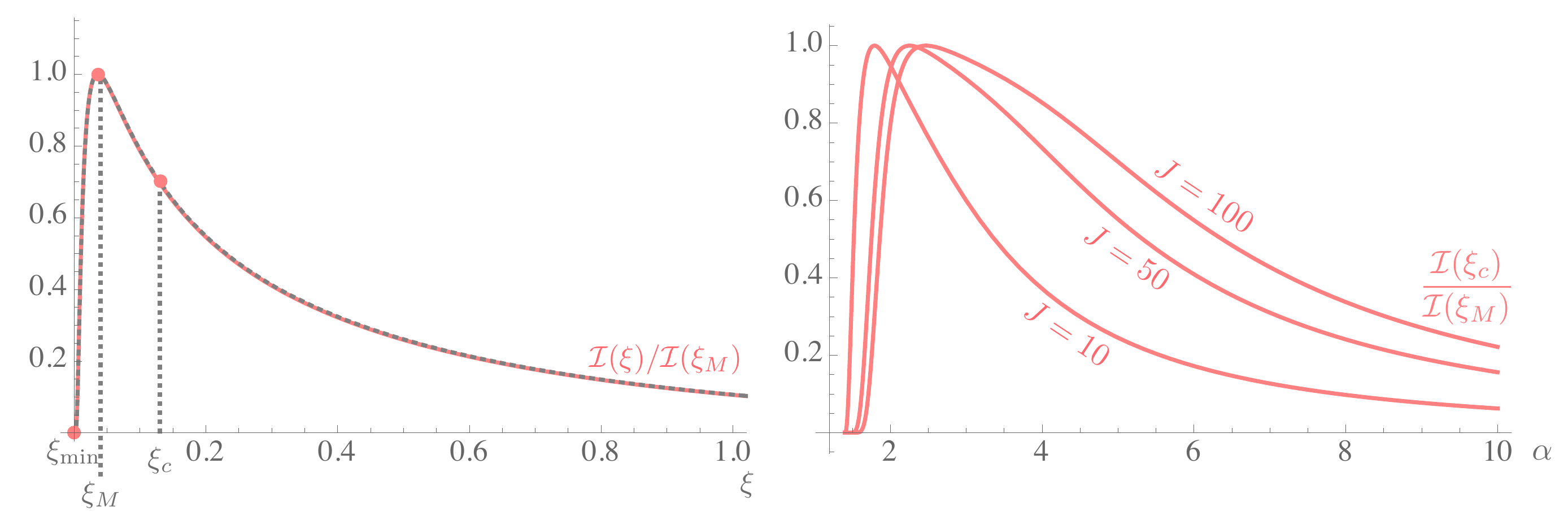}
\caption{Left: $\mathcal{I}(\xi)$ in units of $\mathcal{I}(\xi_\text{M})$ for $J=100$, $|\Lambda|=M_{UV}^4$ , $v=M_{UV}/5$ and $\frac{\Delta v }{v}=0.1$ implying $\xi_{\min}=4\times 10^{-4}$ and $\xi_M=0.036$. Continuous pink line corresponds to the exact result of Eq. \eqref{eq:Integral} while the dashed grey line represents the approximation of Eq. \eqref{eq:Iapprox}. Right: $\frac{\mathcal{I}(\xi_c)}{\mathcal{I}(\xi_M)}$ as a function of the hierarchy parameter $\alpha$. }
\label{fig:IPlot}
\end{figure}

Comparing $\mathcal{I}(\xi,\alpha)$ for $\xi_{\min}$, $\xi_{\text{M}}$ and $\xi_\text{c}$  one concludes that 
\be
 \mathcal{I}(\xi_{\min}) \ll \mathcal{I}(\xi_{\text{c}}) <\mathcal{I}(\xi_{\text{M}} )
\ee 
for all $\alpha$, as illustrated in Fig. \ref{fig:IPlot}.

With these estimates of the intersection volume in hand, we can now turn them into a bound on the brane charges by demanding that $\mathcal{V}_\text{cell}=q^J \le \mathcal{V}_\cap(\xi,\alpha)$. The most favourable case corresponds to $\xi=\xi_M$  with $\alpha^4 \bar{\lambda}\gg1$, as this is the one where $\mathcal{V}_\cap$ is maximised, where one finds
\be
q\le[ \mathcal{V}_\cap(\xi_\text{M},\alpha^4 \bar{\lambda}\gg1)]^\frac{1}{J}\equiv  q_\text{GL}(\xi_\text{M},\alpha^4 \bar{\lambda}\gg1)\ .
\ee

It is instructive to compare this bound with the charges from the BP mechanism, Eq. \eqref{eq:qCCBound}:
\be
q_\text{GL}(\xi_\text{M},\alpha^4 \bar{\lambda}\gg1)\ \approx q_{BP} \left(\frac{\Delta v}{v}\right)^\frac{2}{J}.
\ee
Assuming that $v=M_P$, which would be the worst case scenario (and one in which $\Lambda> M_P^4$) $\frac{\Delta v}{v} \approx 10^{-16}$ implying that for $J>32$ the two are of the same order of magnitude. Note that for the benchmark point of $J=100$ of \cite{Bousso:2000xa} we find that $q_\text{GL} \approx q_{BP}/2$ and indication that a simultaneous solution to the CC and EW hierarchy problems can be found in the presence of a mild hierarchy between $v$ and $\Lambda^{1/4}$ at the expense of introducing a weak coupling  between the four-forms and the Higgs scalar.

We note that the steep fall-off of $\mathcal{I}(\xi)$ as $\xi\rightarrow \xi_{\min}$, will affect the bounds on the charges. In this limit, the intersection of the constraints takes place close to the north pole, $\mathcal{V}_\cap$ will be significantly smaller and one can show that
\be
q_\text{GL}(\xi_{\min})\ \approx q_{BP} \left(\frac{\Delta v}{v}\right)^{1-\frac{1}{J}}\ \ll \ q_\text{GL}(\xi_\text{M}) \ , \ q_\text{BP}\ .
\ee
This highlights the crucial role of $\xi$ (or equivalently of the coupling $g$) in this model: decreasing $\xi$  with respect to the optimal value $\xi_M$ and the upper bound of the charge becomes stricter by a factor of $\frac{\Delta v}{v}\ll1$, decrease it any further and there will be no vacua with both the correct CC and EW scales. If instead one considers $\xi>\xi_M$ the effects are milder, as can be seen from the slow fall off to the right of the maximum in Fig. \ref{fig:IPlot}, and the bounds on the charges will be less stringent when compared to the $\xi=\xi_{\min}$ case.

\section{Discussion}\label{sec:conclusions} 

In this paper we investigated whether selfish/Goldilocks Higgs models could be extended to accomodate vacua with both the right Higgs mass and cosmological constant without imposing either scale as input parameters. We found that by increasing the number of four-forms that couple to the scalar, the phenomenology of these models was significantly enriched.
In particular, a simple extension of the mechanism to two coupled four-forms automatically allows, through the linear structure of the constraints on flux space, to find vacua with the correct electroweak scale, regardless of the coupling strength and of the  brane charges. Furthermore in the presence of a large number of four-forms vacua with the correct Higgs mass and cosmological constant exist, provided the scalar-four-form coupling is larger than the lower bound of Eq. \eqref{eq:xi_min}. We derived the corresponding bounds on the brane charges, the only dimensionful free parameter of the model, and have shown how these depend on $g$ and on the ratio between the bare mass scales associated with the CC and EW scales. For the most favourable setup, when $\xi=\xi_M$, these charges are within an $\mathcal{O}(1)$ factor of the values necessary to solve the CC problem alone through the BP mechanism. We therefore conclude that by a simple extension of the well known BP landscape one can turn the fine tuning problems associated with the Higgs mass and with the CC into the issue of finding the environment that is "just right" in a rich flux landscape. This can be achieved without inserting either physical scale as input to the model.

\section*{Acknowledgments}

We would like to thank Alexander Westphal for comments on the draft.

%

\appendix\section{Numerical method and results}

In this appendix we lay out the details of the numerical method employed to validate the analytical estimates for the intersection volumes presented in Sec. \ref{sec:J4Forms}.\\
The procedure relies on a Montecarlo method for the random generation of points in a $J$-dimensional space: generating points in a uniform way in a region of space, by counting the number of those points that satisfy the desired constraints (that is, to be both inside the shell and between the planes) one can estimate the volume of the intersection region. In particular, let us call:
\begin{itemize}
\item $\mathcal{V}_\cap$ and $\mathcal{V}_\text{cube}$ the volume of the intersection region and volume of the cube containing the spheres respectively;
\item $N_\cap$ and $N_{cube}$ the number of points satisfying the conditions and the total number of points.
\end{itemize}
If the points are generated at random one can estimate $\frac{\mathcal{V}_\cap}{\mathcal{V}_\text{cube}}=\frac{N_\cap}{N_{cube}}$, with precision increasing with the total number of points generated. The simplest approach would be to uniformly generate points inside a  $J$-dimensional cube with sides tangent to the outer sphere, however this method becomes computationally inefficient for high values of $J$. Indeed as $J$ grows, the $\mathcal{V}_\text{shell}/\mathcal{V}_\text{cube}$, decreases practically exponentially \footnote{Furthermore, we are interested in those points inside the shell that are also between the planes, increasing again the number of rejected points.}
\begin{align}
& \mathcal{V}_\text{cube} = \left( 2 (R + \Delta R ) \right) ^J, \\
& \mathcal{V}_\text{shell} = \frac{\pi ^{J/2}}{\Gamma \left( \frac{J}{2} + 1 \right)} \left( (R + \Delta R) ^J - R^J \right),
\end{align}
so that $\frac{\mathcal{V}_\text{shell}}{\mathcal{V}_\text{cube}} \ll 1$ and the statistics quickly deteriorates.\\
In order to overcome this numerical obstacle  we employ the Marsaglia algorithm to draw points uniformly directly inside the shell. By counting the points that lie between the planes one can estimate $\mathcal{V}_\cap$, in an efficient manner.\\
It turns out that, if $r$ is a number generated uniformly at random in the interval $[0,1]$ and $\bm{n}$ is a point selected uniformly at random from the unit $(J-1)$-sphere, then $r^{1/J} \bm{n}$ is uniformly distributed within the unit $J$-ball. In order to generate the points we follow these steps:
\begin{enumerate}
\item Generate the radius in $\left[R, R + \Delta R \right]$: draw uniformly a number in $\left[ \left( \frac{R}{R + \Delta R} \right) ^J, 1 \right]$, exponentiate it to the power of $1/J$ and multiply it by $R + \Delta R$. Call this value $\mathcal{R}$.
\item Generate the point $P$ on a $(J-1)$-sphere: draw $J$ coordinates $x_i$, with $i = 1,...,J$ using a Gaussian distribution with mean equal to zero and arbitrary variance\footnote{While the mean of the Gaussian distribution must be zero, the variance can be arbitrarily chosen, and in particular, we choose it to be $1$.}.
Using these coordinates find the distance from $P$ to the origin, and call it $d$; this means that $P$ has been uniformly generated on a sphere with radius $d$.
\item Move the point to the radius sphere of radius $\mathcal{R}$, found in the first step: multiply all the coordinates $x_i$'s by the quantity $\mathcal{R}/d$. Now the point $P$ is on the desired sphere.
\end{enumerate}
Once the point has been generated within the shell region, using this procedure, it is sufficient to evaluate if it also between the planes.\\

For the numerical method we cannot use the parameters found in the theoretical computations, $\Delta R \approx 10^{-121} M_P^2$ and $\Delta v^2 \approx 10^{-34} M_P^2$, as these are too small and would require a generation of $\sim 10^{33}$ points just to find one intersection point.  In order to overcome this numerical issue we rescaled parameters that allow us to gather significant statistics while maintaining the total number of points below $2\cdot 10^9$, but preserving the hierarchy
\be
\label{eq:hierarchy deltaR-deltav-R}
\Delta R \ll \frac{\Delta v^2}{\sqrt{2\xi}} \ll R
\ee
in order to be able to apply the analytic approximations described in Sec. \ref{sec:Vint}.\\
The chosen values for the parameter, in units of $M_{UV}^2$ (apart from $\alpha$ which is dimensionless), are:
\begin{itemize}
\item $R \approx 1.5$;
\item $\Delta v^2 = 0.02$;
\item $\Delta R = 0.001$;
\item $\alpha = 2$.
\end{itemize}

\begin{figure}
\centering
\subfloat[][\emph{Comparison in $\xi = \xi_{min}$ configuration.}]
{\includegraphics[width=0.85\textwidth]{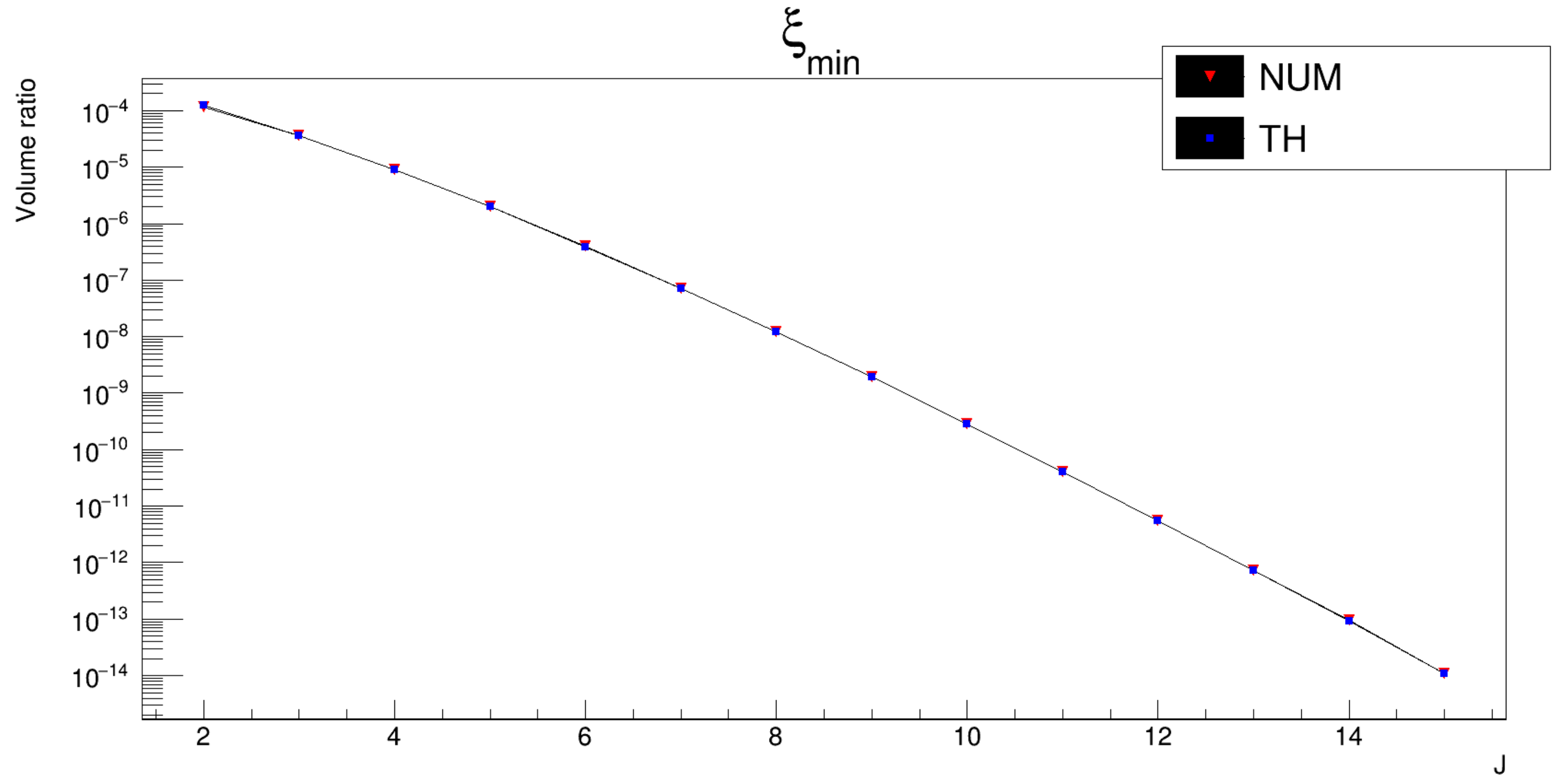}
\label{subfig:gmin}} \\
\subfloat[][\emph{Comparison in $\xi = \xi_{M}$ configuration.}]
{\includegraphics[width=0.85\textwidth]{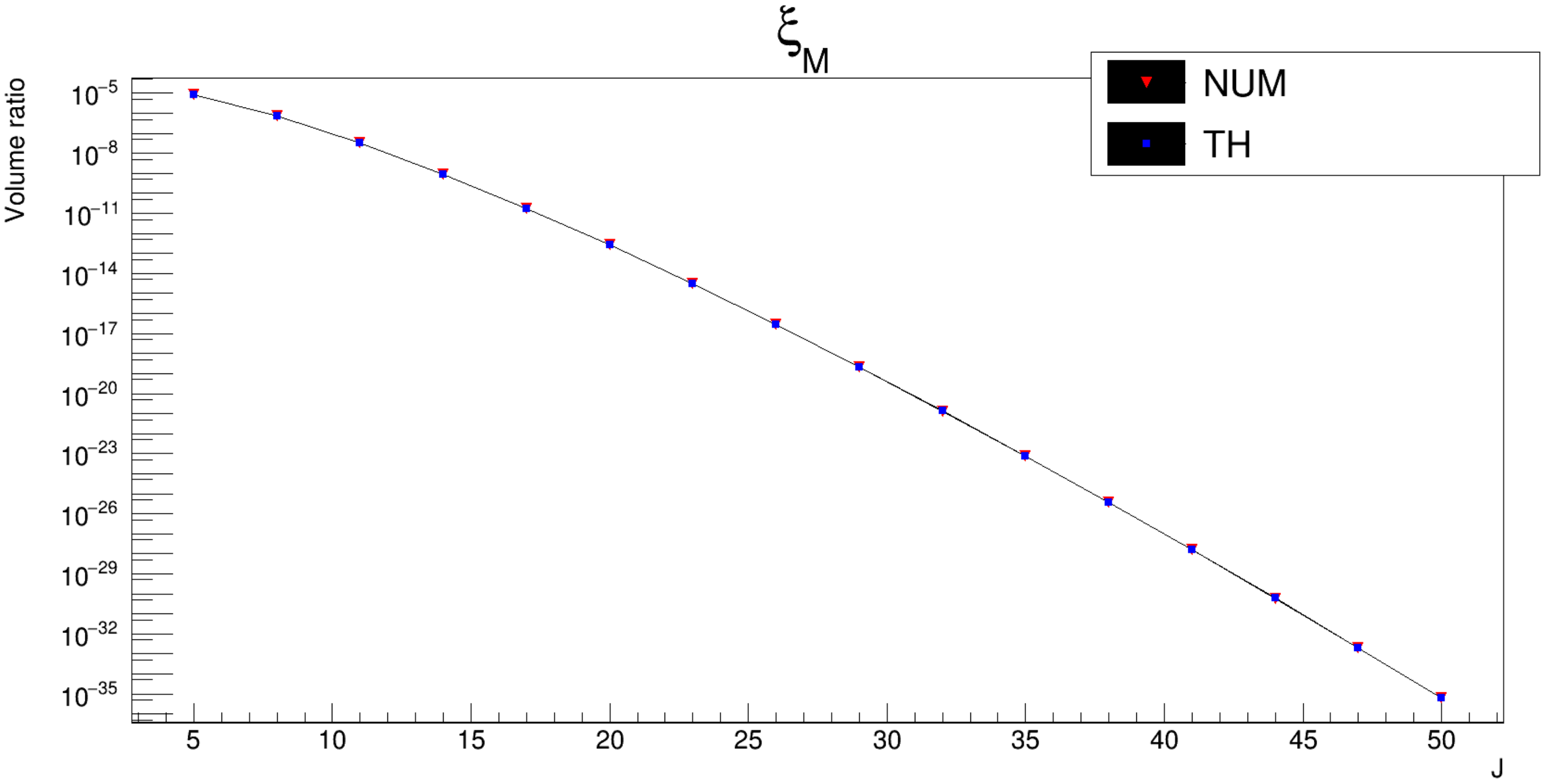}
\label{subfig:gmax}}
\caption{Comparison between numerical and analytical results for $\xi = \xi_{\min}$ (top) $\xi = \xi_{\text{M}}$ (bottom). Red points denote the numerical results while the blue ones denote the theoretical results. The graphs are essentially overlapped showing a good agreement between the theoretical and numerical approaches.}
\label{fig:graphs}
\end{figure}

These choices are compatible with Eq. \eqref{eq:hierarchy deltaR-deltav-R} and are driven mostly by computational convenience.\\Once the parameters have been chosen we can compare the numerical result $\frac{N_\cap}{N_\text{cube}}$ with the analytical ones $\frac{\mathcal{V}_\cap}{\mathcal{V}_\text{cube}}$ found with the following formula\footnote{Even if in this notation we write $V_{cube}$ and $N_{cube}$, we do not actually refer to the cube as random generation points region, but rather to the one illustred previously, \emph{i.e.} the shell.}:
\begin{equation}
\label{eq:volume ratio}
\frac{\mathcal{V}_\cap}{\mathcal{V}_\text{cube}} = \frac{\Delta R \pi^\frac{J-1}{2}}{2^{J-1} R \, \Gamma \left( \frac{J-1}{2} \right)} \int \, du (1-u^2)^\frac{J-3}{2}.
\end{equation}
The integration limits are different for the two configurations and are given by Eqs. \eqref{eq:umin} and \eqref{eq:umax}. 
For $\xi = \xi_{min}$ the  statistics quickly deteriorates with the number of dimensions, forcing us to stop at $J=15$ (figure [\ref{subfig:gmin}]), while for $\xi = \xi_M$ the fact that the planes are closer to the equator makes the statistic to be slightly better so that we can push $J$ up to $50$ (figure [\ref{subfig:gmax}]).
In both the cases the numerical results agree with analytical estimates presented in the main text.

\end{document}